\documentclass[preprint,aps,prl,preprintnumbers,amsmath,amssymb,superscriptaddress,showpacs,floatfix]{revtex4}
\usepackage{epsfig}
\usepackage{dcolumn}
\usepackage{bm}
\usepackage{amssymb}
\usepackage[bookmarks=false]{hyperref}

\begin{document}

\title{The Heavy Quark Free Energy in QCD and in Gauge Theories with Gravity Duals}

\author{Jorge Noronha}
\affiliation{Department of
Physics, Columbia University, 538 West 120$^{th}$ Street, New York,
NY 10027, USA}

\begin{abstract}

We show that the Polyakov loop approaches its asymptotic value at high temperatures {\it from below} in any strongly-coupled gauge theory at large $N_c$ with a dual supergravity description. In these theories the Polyakov loop can be used to define a quantity that behaves as a free energy of a heavy quark in the medium. This should be compared to the recent lattice results in pure glue $SU(3)$ theory at high temperatures where the Polyakov loop approaches its non-interacting limit {\it from above}. We show that the ``heavy quark free energy" obtained from the renormalized loop computed on the lattice does not behave as a true thermodynamic free energy. We conjecture that this should always be the case in asymptotically free gauge theories.  

\end{abstract}

\maketitle


$SU(N_c)$ gauge theories considerably simplify when the number of colors, $N_c$, is large \cite{largeNc}. The deconfinement phase transition in this limit is predicted to be a strong first order transition where the latent heat $\sim N_c^2$ and, in fact, this has been shown to be true in lattice calculations performed in pure gauge theories with $N_c \geq 3$ \cite{largeNclattice}. In this paper I show that when $N_c\to \infty$ the expectation value of the (regularized) Polyakov loop, which is the order parameter \cite{polyakov} for the deconfining phase transition in $SU(N_c)$ gauge theories without dynamical fermions, becomes a monotonic function of the temperature in the deconfined phase of gauge theories with supergravity duals. 

The order parameter for deconfinement in an $SU(N_c)$ theory without dynamical fermions is defined in terms of the path ordered Polyakov loop \cite{polyakov} 
\begin{equation}
\ \bold{L} (\vec{x})=P \,e^{i\,g\int_{0}^{1/T} \hat{A}_{0}(\vec{x},\tau)d\tau}
\label{polyakovdef}
\end{equation}
where $\hat{A}^{0}$ is the non-Abelian gauge field operator and $g$ is the coupling. This operator becomes gauge invariant after performing the trace over the fundamental representation of $SU(N_c)$, denoted here as ${\rm tr}$. Note that ${\rm tr\,}\bold{L}$ transforms as a field of charge one under the global $Z(N_c)$ symmetry, i.e., ${\rm tr\,}\bold{L} \to e^{i 2\pi a/N_c}{\rm tr\,}\bold{L}$ where $a=0,\ldots,N_c-1$. The magnitude of the expectation value of this operator after the (normalized) trace is denoted here as $\ell \equiv |\langle {\rm tr\,}\bold{L}\rangle|/N_c$. Below $T_c$ the system is $Z(N_c)$ symmetric, which implies that $\langle {\rm tr\,}\bold{L} \rangle = 0$. Above $T_c$ this global symmetry is spontaneously broken, $\ell(T)\neq 0$, and the system lands in one of the possible $Z(N_c)$ vacua.  

Wilson lines are renormalizable operators \cite{wilson}. The expectation value of a Wilson line has a mass divergence that depends on the length of the loop \cite{wilson} and for a thermal Wilson line the length is $1/T$ and $\ell \sim \exp(-m^{div}/T)$, where $m^{div}$ is a divergent mass that is linear in the UV cutoff. On a lattice with spacing $a$ one would have $m^{div}\sim 1/a$. It is expected that $m^{div}>0$ and, thus, the bare quantity computed on the lattice vanishes in the continuum limit \cite{Dumitru:2003hp}. The wave function renormalization of the loop is given in terms of $\mathcal{Z}=\exp(-m^{div}/T)$, which defines then the renormalized quantity $\bold{L}_R = \bold{L}/\mathcal{Z}$ \cite{Dumitru:2003hp} (we only consider smooth loops here). We shall assume that all the quantities discussed here have been properly renormalized so that we can drop the subscript $R$ in the following. Moreover, note that $\ell$ is a positive-definite number that vanishes below $T_c$ in pure glue gauge theories.

The thermal average of the 2-point function $\mathcal{C}(r,T)\equiv\langle {\rm tr}\bold{L}^{\dagger}(r)\,{\rm tr}\bold{L}(0)\rangle / N_c^2$ is usually associated with the free energy of an infinitely heavy $Q\bar{Q}$ pair separated by a distance $r$ in the medium \cite{McLerran:1980pk}. It is appropriate at this point to re-examine this argument when $N_c$ is large. We use the standard definition for the thermal expectation value of a given operator $\mathcal{B}$ \cite{kapusta}
\begin{equation}
\langle \mathcal{B} \rangle = \frac{{\rm Tr }\,\mathcal{B}\,e^{-\mathcal{H}/T}}{{\rm Tr }\,e^{-\mathcal{H}/T}}
\label{defaverage}
\end{equation} 
where $\mathcal{H}$ is the Hamiltonian of the system and ${\rm Tr}$ denotes the trace operation defined in terms of a path integral over the fields. 

The free energy of a system composed by gluons and a $Q\bar{Q}$ pair is    
\begin{equation}
\ F(\vec{r}_1,\vec{r}_2,T)=-T\,\ln \sum_{|s\rangle} \langle s|e^{-\mathcal{H}/T}|s \rangle
\label{freeenergyQQbar}
\end{equation}
where $\mathcal{H}$ is the Hamiltonian of the whole system (including the quarks), and the sum is over all the states $|s\rangle$ with a heavy quark at $\vec{r}_1$ and an anti-heavy quark at $\vec{r}_2$ (with $r\equiv |\vec{r}_1-\vec{r}_2|$). After introducing creation and annihilation operators for the quarks and anti-quarks and regularizing the loops, one can define a renormalized $Q\bar{Q}$ interaction energy
\begin{equation}
\Delta F_{Q\bar{Q}}(r,T)\equiv F(r,T)-F^{g}(T) = -T\ln \mathcal{C}(r,T)
\label{freeenergyQQbar1}
\end{equation}
in terms of the difference between the free energy of the system with the quarks, $F(r,T)$, and the free energy of pure glue at finite $T$, $F^{g}(T)$. 
      
The total free energy of the pure glue medium can be written as \cite{Pisarski:1983db} 
\begin{equation}
\ F^{g}(T)=N_c^2 F^{g}_2(T)+F^{g}_0(T)+\mathcal{O}(1/N_c^2)
\label{defineFglue}
\end{equation} 
where the coefficients $F^g_i=F^g_i(\lambda,T)$ are only functions of $T$ and the t' Hooft coupling $\lambda_{YM}= g_{YM}^2 N_c$. A similar expansion must also hold for the system which includes the effects from heavy quark probes
\begin{equation}
\ F(r,T)=N_c^2 F_2(T)+F_0(r,T)+\mathcal{O}(1/N_c^2)
\label{defineFglueplusquark}
\end{equation} 
where again $F_i=F_i(\lambda,T)$ but, in general, $F_i$ and $F^g_i$ are not identical. However, note that $F_2=F_2^g$ since at large $N_c$ the contribution from the quarks cannot enter at that order. Therefore, one can see that 
\begin{equation}
\lim_{N_c \to \infty}\Delta F_{Q\bar{Q}}(r,T)= F_0(r,T)-F_0^{g}(T)\,.
\label{freeenergydifference}
\end{equation}
All the coefficients in the series in Eqs.\ (\ref{defineFglue}) and (\ref{defineFglueplusquark}) define a well-defined free energy order by order in $N_c$, i.e, $F_i$ is the contribution of order $N_c^i$ to the total free energy, which must behave as true thermodynamic free energy \cite{landau} as well. This argument is certainly valid when $N_c\to \infty$ and all the coefficients in the series above can be unambiguously defined. 

Note that $\ell^2(T) = \mathcal{C}(r \to \infty,T)$ and with this relation one can obtain a quantity that is associated with a single heavy quark in the medium via 
\begin{equation}
\ F_{Q}(T) \equiv - T \ln \ell(T). 
\label{defineheavyquarksinngle}
\end{equation}    

The standard physical interpretation is that difference of free energies $\Delta F^R_{Q\bar{Q}}(r,T)$ represents the free energy contribution of the heavy $Q\bar{Q}$ pair to the medium \cite{McLerran:1980pk}. In general, the quantity defined in Eq.\ (\ref{freeenergyQQbar1}) may not behave like a free energy in the thermodynamic sense. The difference between free energies is not guaranteed to be a free energy since the static color-electric field sourced by the $Q\bar{Q}$ pair can considerably change the energy of the system. For instance, in a confining theory the energy of the system below $T_c$ is not a well defined quantity if the quarks are infinitely distant from each other. 

Let us now see under what conditions $F_{Q}(T)$ (and hence $\Delta F_{Q\bar{Q}}(r,T)$) can be associated with a free energy. The derivative of $\ell$ with respect to $T$ is
\begin{equation}
\frac{d\ell(T)}{dT} = \frac{\ell(T)}{2T^2}\,\left(U_0(r\to \infty,T)-U_0^{g}(T)\right).
\label{heavyquarkfree}
\end{equation} 
where $U_0(r,T)=F_0(r,T)-T\,dF_0(r,T)/dT$ is the internal energy of the system that contains a quark while $U_0^{g}(T)=F_0^g(T)-T\,dF_0^g(T)/dT$ is the corresponding quantity obtained in the absence of a quark. As argued above, each of these internal energies should display the expected thermodynamic behavior when defined in terms of the free energy \cite{landau}. 

Since in a confining theory $\ell$ is zero at $T<T_c$ and positive otherwise, we see that immediately above $T_c$ the derivative $d\ell/dT > 0$. This means that the difference in internal energies in Eq.\ (\ref{heavyquarkfree}) is positive at $T \to T_c^{+}$. In homogenous systems in equilibrium the internal energy can only increase with $T$ since the specific heat must be non-negative \cite{landau}. However, this does not imply that $d\ell/dT$ should be always positive. In fact, one can imagine that because of the quark the specific heat of the system in this case can increase at a different rate and at higher temperatures the difference in internal energies in the equation above can vanish and then become negative. If this happens, $F_{Q}(T)$ cannot be a free energy because its corresponding ``internal energy" $U_Q(T)\equiv U_0(r\to \infty,T)-U_0^{g}(T)$ would have negative specific heat $C_Q(T)=dU_{Q}(T)/dT < 0$. On the other hand, if the derivative of the Polyakov loop remains positive at any finite $T>T_c$ then $F_Q(T)$ acquires a true thermodynamic meaning, $C_Q(T)>0$, and $d\ell(T)/dT \to 0^{+}$ when $T\to \infty$ where there is usually an UV fixed point.        

Perturbative QCD calculations in the hard-thermal-loop approximation performed using a running coupling constant \cite{Gava:1981qd,Burnier:2009bk} suggest that $\ell$ should approach one from above. Recently, lattice simulations \cite{Kaczmarek:2002mc,Gupta:2007ax} in $SU(3)$ pure glue showed that $\ell$ increases very rapidly after its jump at $T_c$ and, in fact, it overshoots one around $3T_c$ and keeps increasing until it reaches a maximum near $10T_c$ \cite{Gupta:2007ax} where it starts to decrease with increasing temperatures towards a constant value that is close to one. According to the discussion above, the fact that in these calculations $d\ell/dT$ changes sign shows that $F_Q(T)$ (or equivalently, $\Delta F_{Q\bar{Q}}(r,T)$) cannot be identified with a quark free energy in the QCD plasma. In this case, the associated ``heavy quark specific heat" $c_Q <0$. The thermal Wilson line is just the trace of the propagator of a quark which, because of its infinite mass, moves in a straight line in imaginary time \cite{Pisarski:2002ji}. This quantity obviously depends on $T$ but this dependence need not be that of a thermodynamic function such as the free energy. 

This issue regarding the physical interpretation of the Polyakov loop correlator was discussed, for instance, in \cite{Pisarski:2002ji} while in \cite{Dumitru:2005ng} it was shown that in the presence of a finite baryonic chemical potential $F_Q$ is expected to increase with $\mu$, which clearly shows that this quantity is not a true free energy. It is more appropriate perhaps to call $F_Q(T)$ simply the heavy quark interaction energy. 

The dominant finite $T$ contribution to the binding energies of very tightly bound pairs, such as the $\Upsilon^{1S}$, comes from $F_Q(T)$ \cite{agnes,adrian,noronhaheavyquark}. If in QCD this quantity is not a true free energy, then the usual ``entropy subtraction" performed in quarkonia binding energy studies \cite{agnes} in pure glue theories may be less justified. This will affect quantitative analyses of quarkonia melting, which is an important signature for deconfinement \cite{agnes}.        

It is natural to think that asymptotic freedom may be the reason for this non-monotonic behavior with $T$ displayed by the Polyakov loop. In fact, the hard-thermal-loop calculation of $\ell$ performed at leading order gives $\ell \sim 1 +\lambda_{YM} (m_D/T)$, where $m_D \sim \sqrt{\lambda_{YM}} \,T$ is the Debye mass. For a fixed value of the coupling $m_D/T$ is a constant and $\ell$ is a number larger than one. Clearly, once the coupling is allowed to run $d\ell/dT <0$ because $d\lambda_{YM}/dT <0$. Thus, while confinement implies that immediately above $T_c$ the derivative of $\ell$ is positive, at high $T$ Debye screening and asymptotic freedom impose that $d\ell/dT$ must be negative. These general arguments lead us to conjecture that in confining gauge theories with a trivial UV fixed point there must be a finite value of temperature, say $T^*$, where $d\ell(T^*)/dT=0$. The exact value of $T^*$ may not be obtained perturbatively unless it is sufficiently larger than $T_c$. As discussed above, in this class of theories $F_Q(T)$ cannot be a true thermodynamic free energy. In fact, since $d\ell/dT <0$ at high $T$ this implies that the specific heat of the gluon medium decreases (by a small amount) due to the presence of the heavy quark.

Since the thermodynamic properties of pure gauge theories computed on the lattice do not change significantly when $N_c\geq 3$ \cite{largeNclattice}, one should expect that $\ell$ is not always a monotonic function of $T$ in QCD at large $N_c$. The published lattice data for the Polyakok loop in full $SU(3)$ QCD with dynamic fermions \cite{Bazavov:2009zn} is still restricted to a much narrower temperature range than the calculations performed in pure glue \cite{Gupta:2007ax}. The general arguments discussed here indicate that in full QCD $\ell$ is not be a monotonic function of $T$, which means that $F_Q$ should not be associated with a free energy. It would be interesting to see if $T^*$ in this case shifts towards the transition region or remains close to the value in pure glue $\sim 10T_c$ \cite{Gupta:2007ax}. However, one should also expect that the exact location of $T^*$ could depend on the regularization scheme used to obtain to compute $\ell$ on the lattice.         

What happens in gauge theories with nontrivial UV fixed points? Since these theories are not asymptotically free, one may expect that the general argument presented above does not apply and in this case $F_Q(T)$ may behave as a free energy. The gauge/string duality \cite{maldacena,witten} provides a way to study the behavior of the Polyakov loop in this class of theories. We shall show in the following that in confining gauge theories with a supergravity-like dual the Polyakov loop always approaches its asymptotic value at high $T$ from below.   

Let us first recall how the deconfinement phase transition takes place in these theories \cite{witten}. The general idea is that the deconfined phase of the 4d gauge theory is dual to a theory of gravity defined in a higher-dimensional geometry with a black brane whose Hawking temperature equals the temperature of the plasma. The free energy of the system above $T_c$ equals the free energy of the black brane and is of order $N_c^2$ (because the effective gravity coupling $\sim 1/N_c^2$) while at lower temperatures another solution of the supergravity equations that does not have a horizon has a smaller free energy, which is of order $N_c^0$ after the inclusion of quantum corrections. The deconfinement phase transition in this approach is predicted to be a strong first order transition with latent heat $\sim N_c^2$, which agrees with lattice calculations of pure glue gauge theories at large $N_c$ \cite{largeNclattice}.        

We shall assume in the following that $N_c \to \infty$ and the radius of the background spacetime $R$ is much larger than the string length $\sqrt{\alpha'}$ (the supergravity limit). For instance, in the planar limit of $N=4$ Supersymmetric Yang-Mills (SYM) theory the t'Hooft coupling is $\lambda_{SYM}=R^4/\alpha'^2$ \cite{maldacena,witten}. When a theory admits a supergravity description several simplifications occur. For example, because the classical supergravity action goes like $\sim N_c^2$, the Euclidean path integral is dominated by the spacetime that is a solution of the classical gravitational equations of motion and it becomes $\sim \exp (-S_{sugra})$ with the on-shell action being $S_{sugra}\sim N_c^2 \int d^{10}x \,\sqrt{-G} \left(\mathcal{R}+\ldots\right)$ while $\ldots$ denote the contribution from the other supergravity modes. 

An infinitely massive quark in the 4d non-Abelian gauge theory at finite $T$ is dual to a string in the bulk that hangs down from a probe D-brane at infinity and reaches the black hole horizon in the bulk spacetime \cite{Maldacena:1998im}. The holographic representation of $\ell$ is simply $\ell(T)=\exp(-S_{NG}(\mathcal{D}))$ where $S_{NG}$ is the classical Nambu-Goto action (in Euclidean time), which is proportional to $R^2/\alpha'$, and $\mathcal{D}$ is the worldsheet that is a solution of the equations of motion describing a single string in the bulk (see \cite{Bak:2007fk,Andreev:2009zk,Noronha:2009ud}). Here the string is indeed a real probe that does not backreact on the geometry and the full partition function factorizes $Z_{Q}(T)/ Z(T)= \exp(-S_{NG})$ (we are assuming that the divergence in the string action that appears near the boundary has been already properly removed). 

Similar arguments to those used in Eqs.\ (\ref{defineFglue}-\ref{freeenergydifference}) also hold in this case but the important difference here is that $F^{g}(T)=N_c^2 F^{g}_2(T)$ when $N_c\to \infty$ and $R^2/\alpha'\to \infty$, i.e, the other coefficients in the $N_c$ expansion of the theory without the quark probe vanish in this classical approximation. Also, $F^{g}_2$ should not depend on $R^2/\alpha'$ in this limit \cite{comment1}. This means that in the classical supergravity approximation $F_{Q}(T)= F_0(r\to\infty,T)/2$. Thus, in the supergravity limit $F_Q(T)$ behaves as a free energy and, according to the general discussion above, $\ell$ should be a monotonic function of $T$. The free energy in this limit is simply given by the Nambu-Goto action via $F_{Q}=T\, S_{NG}$ evaluated using the background spacetime. The specific heat of the plasma increases (by a very small amount) when the heavy quark is added, which is the opposite of what occurs in a pure glue plasma according to the discussion presented earlier involving the lattice data.     

Since screening also takes place at strong coupling \cite{Bak:2007fk}, it seems that the reason why $\ell$ behaves this way in these theories is because they are not asymptotically free. One may argue that holographic methods can provide an approximation for $\ell$ in QCD only when $T < T^*$ since above $T^*$ effects from asymptotic freedom are not negligible (even though the coupling around $T^*$ may not be small). However, when corrections to supergravity of order $R^2/(\alpha'\,N_c^2)$ are present in the bulk the string cannot be seen as probe and it may be possible to show that there is a finite $T^*$ where $d\ell/dT$ vanishes.

Thus, we see that the temperature dependence of $\ell$ in confining theories at large $N_c$ varies with the type of fixed point present in the UV. For theories that are asymptotically free there must be a finite value of $T>T_c$ where $d\ell(T)/dT=0$ while for theories with nontrivial fixed points described by supergravity duals $\ell$ is a monotonic function of $T$ in the deconfined phase. However, we would like to stress again that once corrections to the supergravity background are taken into account $\ell$ may not be a monotonic function of $T$. This could be the case even if this higher order theory is not asymptotically free.

This result for the $T$ dependence of $\ell$ in gauge theories computed in the supergravity approximation can be used to show that strongly-coupled gauge theories at large $N_c$ with nontrivial UV fixed points may display unphysical results depending on the scaling dimension, $\Delta < 4$, of the operator used to deform the original CFT and induce a nontrivial renormalization group flow. The general class of gravity duals we are interested in were studied in detail in \cite{Gubser:2008ny} where it was shown that it is possible to obtain black brane backgrounds with equations of state that are similar to the QCD equation of state obtained on the lattice (see also \cite{Gursoy:2008bu} for a model involving the dilaton). 

The gravity action describes the interactions between a 5d metric (in the Einstein frame) $G_{\mu\nu}$ and a single scalar field $\phi$ \cite{Gubser:2008ny}  
\begin{equation}
\mathcal{A}=\frac{1}{2 k_5^2} \int d^5x\sqrt{-G}\left[\mathcal{R}-\frac{(\partial \phi)^2}{2}-V(\phi)\right].
\label{dilatonaction}
\end{equation}  
where $k_5^2 \sim 1/N_c^2$. Conformal invariance in the UV can be obtained by imposing that $\phi\to 0$ at the boundary and
\begin{equation}  
\lim_{\phi\to 0} V(\phi)=-\frac{12}{R^2}+\frac{1}{2 R^2}\Delta(\Delta-4)\phi^2+\mathcal{O}(\phi^4),
\label{conditionboudnary}
\end{equation}
where the mass squared of the scalar $m_{\phi}^2 R^2=\Delta(\Delta-4)\geq -4$ and, according to the Breitenlohner-Freedman bound \cite{BFbound}, $1\leq \Delta < 4$. Eq.\ (\ref{conditionboudnary}) implies that the spacetime will be asymptotically AdS$_5$ with radius $R$. The deviations from conformal behavior only become significant deep down in the geometry when $u$ is small. The boundary of the bulk spacetime is located at $u \to \infty$. The Ans\"atze for the Einsten frame metric and the scalar are 
\begin{equation}
\ ds^2 = a^2(u)(-f(u) dt^2+d\vec{x}^2)+\frac{du^2}{b^2(u)\,f(u)},\,\, \phi=\phi(u)\,.
\label{metric}
\end{equation} 
Finite temperature effects are included by considering solutions that display a horizon at $u=u_h$ where $f(u_h)=0$. At temperatures $T/\Lambda \gg 1$, where $\Lambda$ is the energy scale associated with the scalar, one should be close to the $AdS_5$ solution and in this case $\phi_h \equiv \phi(u_h) \ll 1$. 

A consistent high $T$ expansion of several transport coefficients in gravity duals with the action $(\ref{dilatonaction})$ and $2<\Delta<4$ was developed in \cite{Cherman:2009kf} and a generalization of these results when $1\leq \Delta < 4$ was done in \cite{Yarom:2009mw}. At temperatures where $\phi_h\ll 1$ the corrections to the CFT values (the $T\to \infty$ limit) start at $\phi_h^2$ where 
\begin{equation}
\phi_h = \frac{\Gamma(\Delta/4)^2}{(\Delta-2)\Gamma(\Delta/2-1)}\left(\frac{\Lambda}{\pi T}\right)^{4-\Delta}\,.
\end{equation}

Following the results presented in \cite{Noronha:2009ud}, one can see that Nambu-Goto action or $F_Q(T)$ is given by 
\begin{equation}
\ F_Q(T)=T_s \, \int_{u_h}^{\infty}du\,q(\phi(u))\frac{a(u)}{b(u)}
\label{FQ}  
\end{equation}
apart from a temperature independent term used to remove the divergence at $u\to \infty$. Here $T_s = 1/(2\pi \alpha')$ is the fundamental string tension and $q(\phi)$ is a function that describes how the fundamental string couples to the scalar. In the supergravity approximation to type IIB string theory, assuming that $\phi$ is the dilaton this coupling function in the Nambu-Goto action would be equal to $\exp(\phi/2)$. $\mathcal{N}=4$ SYM is conformal invariant and in the supergravity approximation the dilaton is a constant. In this case the gauge coupling $\lambda_{SYM}$ does not run and the expression for the free energy is $F_{Q} = -\sqrt{\lambda_{SYM}}\,T/2$ at strong coupling (the linear dependence with $T$ follows from conformal invariance). Note that in exactly conformal theories such as $\mathcal{N}=4$ SYM the internal energy always vanishes, i.e, $U_Q=0$ to all orders in the t'Hooft coupling.  

In the case of the effective model in Eq.\ (\ref{dilatonaction}) the function $q(\phi)$ can be in principle an arbitrary function of $\phi$. However, we shall show in the following that it is possible to obtain constraints on $\Delta$ and $q(\phi)$ by imposing that $F_{Q}$ in Eq.\ (\ref{FQ}) behaves as a free energy (or, equivalently, that $\ell$ is a monotonic function of $T$) in gauge theories dual to supergravity backgrounds that have a nontrivial fixed point in the UV. These constraints are derived using the high $T$ expansion of $U_Q$ and $C_Q$ computed from Eq. (\ref{FQ}). Since in these gravity duals
\begin{equation}
\frac{d\ell}{dT}=\frac{\ell}{2T^2}U_{Q}(T)
\label{newdldT}    
\end{equation}
and $U_Q(T\sim T_c)>0$, the condition $d\ell/dT \geq 0$ implies that $U_Q$ should remain positive at any $T$ and, thus, $C_Q >0$. While near the transition these quantities can only be computed numerically, one can use the expansion defined in \cite{Cherman:2009kf} to determine their high $T$ behavior analytically. In fact, $\Lambda/T \ll 1$ implies that $\phi_h \ll 1$ and one can expand the coupling function in a power series around $\phi_h \sim 0$, $q(\phi_h)=q(0)+q'(0)\phi_h+q''(0)\phi_h^2/2+\ldots$, to show that the heavy quark entropy is  
\begin{equation}
\ S_Q(T)\equiv -\frac{dF_Q}{dT}=\frac{R^2}{2\alpha'}q(0)\,\sum_{n=0}s_{n}(\Delta)d(\Delta)^n \,\left(\frac{\Lambda}{\pi T}\right)^{(4-\Delta)n}
\label{entropy}    
\end{equation}        
where 
\begin{equation}
\ d(\Delta) = \frac{\Gamma(\Delta/4)^2}{(\Delta-2)\Gamma(\Delta/2-1)}
\label{dfunction}
\end{equation}
and the first coefficients are
\begin{equation}
\ s_0(\Delta)=1\,,\qquad s_1(\Delta)=\frac{q'(0)}{q(0)}\,,
\label{c0c1}
\end{equation}
and 
\begin{equation}
\ s_2(\Delta)=\frac{q''(0)}{2q(0)}+\frac{1}{3\pi}(4-\Delta)(2-\Delta)\tan(\pi \Delta/4) +\frac{\Delta(\Delta-4)}{24}\,.
\label{c2}
\end{equation}
It is easy to show using Eq.\ (\ref{entropy}) that when $\Delta \neq 3$ the internal energy is 
\begin{equation}
\ U_Q(T)=\frac{R^2}{2\alpha'}q(0)\,T\,\sum_{n=1}s_{n}(\Delta)d(\Delta)^n \,\frac{(\Delta-4)n}{(\Delta-4)n+1}\left(\frac{\Lambda}{\pi T}\right)^{(4-\Delta)n}\,,
\label{internale}    
\end{equation}
which leads to the following expression for the specific heat
\begin{equation}
\ C_Q(T)=\frac{R^2}{2\alpha'}q(0)\,\sum_{n=1}s_{n}(\Delta)d(\Delta)^n \,(\Delta-4)n\,\left(\frac{\Lambda}{\pi T}\right)^{(4-\Delta)n}\,.
\label{specificheat}    
\end{equation}
Note that Eq.\ (\ref{internale}) implies that $d\ell/dT\to 0$ when $T/\Lambda\to \infty$. Let us assume that the coupling function is such that $q(0)>0$ and $s_1(\Delta)\neq 0$. Then, by requiring that $C_Q$ is positive at high $T$ one obtains that $s_1(\Delta)<0$ (note that $d(\Delta)>0$ when $1\leq \Delta <4$). However, if $s_1(\Delta)$ is negative then $U_Q$ can only be positive if $\Delta \geq 3$ (the generalization of the formulas above for the case where $\Delta=3$ is straightforward). One can see from Eq.\ (\ref{c0c1}) that this condition implies that $q'(0) <0$. Thus, simple coupling functions that increase exponentially with $\phi$ lead to inconsistencies in the supergravity approximation of gravity theories dual to CFTs deformed by a relevant operator (if $\Delta=4$ the standard exponential coupling is, of course, valid). Exponentially decreasing coupling functions, such as $\exp(-\xi \phi)$ with $\xi >0$, usually appear in the description of D1-branes whose endpoints at the boundary correspond to magnetic monopoles in the plasma. Thus, since the condition $q'(0) <0$ cannot probably be easily understood in terms of fundamental strings, one may want to consider the case where $q'(0)=0$. If $s_1(\Delta)$ is identically zero then $s_2(\Delta)$ must be negative so that $C_Q >0$. This implies that $U_Q$ is positive at large $T$ only if $\Delta>3.5$. Note also that this constraints the value of $q''(0)$ in terms of $\Delta$ via Eq.\ (\ref{c2}). Perhaps a good choice for phenomenological applications would be $q(\phi)=\cosh(\xi \phi)$, where $\xi=\xi(\Delta)$ is chosen to give $s_2(\Delta) <0$. This coupling function becomes exponentially large in the IR and leads to the expected high $T$ behavior for the Polyakov loop when $1\leq \Delta <4$.

It is clear that the allowed values of $\Delta$ become closer to 4 if a large number of higher-order derivatives of $q$ vanish at the boundary. This should be expected since the case where all the derivatives of $q$ are zero at the boundary corresponds to the exactly marginal deformation where $\Delta$=4 and the resulting gauge theory is still a CFT where $U_Q$ and $C_Q$ are identically zero. It would be interesting to see how these consistency conditions change when multiple scalar fields and other supergravity modes are present such as in the supergravity dual of $\mathcal{N}=2^*$ theory \cite{n2star}.

The high $T$ limit of $S_Q$ determines the value of $\ell$ at the nontrivial UV fixed point, i.e, 
\begin{equation}
\lim_{\Lambda/T \to 0}\,\ell(T)=\exp \left(\frac{R^2}{2\alpha'}q(0)\right)  
\label{highTloop}
\end{equation}
up to a $T$ independent integration constant. When the conditions derived above are satisfied, $\ell(T)$ approaches its asymptotic value in Eq.\ (\ref{highTloop}) from below in the class of gravity duals considered here. 

The nontrivial restrictions on the scaling dimension of scalar fields in gravity duals derived here may prove useful in practice to develop consistent deformations of CFT's that can be used as effective models for the strongly-coupled quark-gluon plasma \cite{sQGP} produced in ultrarelativistic heavy ion collisions.     

I am grateful to A.~Dumitru for many helpful comments and M.~Gyulassy, R.~Pisarski, G.~Torrieri, M.~Guimaraes, M.~Panero, D.~Rischke and P.~Petreczky for discussions. This work was supported by the US-DOE Nuclear Science Grant No.\ DE-FG02-93ER40764.

\end{document}